\documentstyle[prb,aps,epsfig]{revtex}
\begin{document}
\draft
\twocolumn[\hsize\textwidth\columnwidth\hsize\csname@twocolumnfalse%
\endcsname

\title{On the Current Carried by `Neutral' Quasiparticles}
\author{Chetan Nayak$^1$, Kirill Shtengel$^2$, Dror Orgad$^1$,
Matthew P.A. Fisher$^3$, and S. M. Girvin$^{3,4}$} 
\address{$^1$Department of Physics and Astronomy,
University of California at Los Angeles, Los Angeles, CA 90095-1547\\
$^2$Department of Physics and Astronomy,
University of California at Irvine, Irvine, CA 92697-4575\\
$^3$Institute for Theoretical Physics, University
of California, Santa Barbara, CA 93106-4030\\
$^4$Department of Physics, Indiana University, Bloomington, IN 47405} 

\date{\today}
\maketitle
\vskip 0.5 cm

\begin{abstract}
The current should be proportional to the momentum in
a Galilean-invariant system of particles of
fixed charge-to-mass ratio, such as an electron liquid in jellium.
However, strongly-interacting electron systems can have phases
characterized by broken symmetry or fractionalization.
Such phases can have neutral excitations which can
presumably carry momentum but not current. In this paper,
we show that there is no contradiction: `neutral' excitations
{\em do} carry current in a Galilean-invariant system
of particles of fixed charge-to-mass ratio. This
is explicitly demonstrated in the context of
spin waves, the Bogoliubov-de Gennes quasiparticles
of a superconductor, the one-dimensional electron gas,
and spin-charge separated systems in $2+1$ dimensions.
We discuss the implications for more realistic systems, which are
not Galilean-invariant.
\end{abstract}

]
\narrowtext

\section{Introduction.}
Conventional wisdom holds that, in
a Galilean-invariant system of particles of
fixed charge-to-mass ratio $e/m$, the local current density
is proportional to the local momentum density,
${\bf J}({\bf x}) = \frac{e}{m} {\bf P}({\bf x})$. The conservation of total
momentum then implies conservation of the total current,
$\frac{d}{dt}{\bf J}=0$. This is a stronger condition than
charge conservation, $\frac{d}{dt}\rho + \nabla\cdot{\bf J}=0$,
since it implies that the real part of the conductivity is given by
$\sigma(\omega) = \frac{n e^2}{m}\, \delta(\omega)$,
where $n$ is the particle density. One might imagine that
this hypothetical situation has some applicability
to extremely clean real systems in which the effects of the lattice
are unimportant because the Fermi surface is far from
any nesting vector and the electron-phonon coupling is very
weak. In such a case, one would be tempted to forget
about impurities and the lattice of ions altogether and focus on the
electrons, which have a fixed charge-to-mass ratio $e/m$.

On the other hand, we have become accustomed to quantum number
fractionalization, particularly
in the context of quasi-one-dimensional materials \cite{Emery79},
the fractional quantum Hall effect \cite{FQHE}, and
theories of high-temperature superconductivity
\cite{highTc,Balents98,Balents99,fractionalization}.
Spin-charge separation is one possible pattern
of quantum number fractionalization. It
leads to charged, spinless quasiparticles -- often called
`holons' -- and neutral, spin-$1/2$ quasiparticles --
often called `spinons'. Conventional wisdom would lead
us to expect that the latter,
being neutral, would carry no current, even when endowed
with non-zero momentum. This is merely the most extreme
and exotic case of a general phenomenon: the low-energy
quasiparticles of a strongly-interacting system need not evince
much resemblance to the underlying electron. This is
true {\em a fortiori} if the low-temperature phase
of the system exhibits fractionalization or
broken symmetry. In particular, there
is no reason why the quasiparticle
charge-to-mass ratio should be $e/m$.
A more familiar, but no less dramatic example
is given by spin waves in a ferromagnet -- neutral spin-$1$ excitations
which carry momentum but, presumably, no current.

Clearly, there is some tension, if not an outright contradiction,
between these two articles of conventional wisdom. The resolution,
which we describe in this paper, is that `neutral' quasiparticles
{\em do} carry current according to
${\bf J}({\bf x}) = \frac{e}{m} {\bf P}({\bf x})$
in a Galilean-invariant system. However, even a small explicit
breaking of Galilean invariance can have drastic consequences
for this relation.  As a result, even a small density
of impurities or a weak periodic potential can result
in a state in which `neutral' quasiparticles carry
momentum but no current and the DC conductivity is zero
rather than infinity.

The current carried by neutral quasiparticles can
be understood as arising from a Doppler shift interaction
between them and the charge carriers. The latter are
always gapless in a Galilean-invariant system,
and they mediate the coupling between the electromagnetic
field and the neutral quasiparticles.
We will illustrate our thesis in a number of different
contexts: spin waves, the Bogoliubov-de Gennes quasiparticles
of a superconductor (which carry momentum but are not charge eigenstates),
the one-dimensional electron gas,
and spin-charge separated systems in $2+1$ dimensions.
Finally, we will comment on our results and their applicability
to realistic systems, which do not have Galilean invariance.

\section{Spin Waves in an Electron Liquid}

As mentioned in the introduction, one might think that
the paradox is already manifest in the context of spin waves
(or other collective excitations) which can carry momentum
but ought not -- if we are to think of them as neutral
excitations -- carry current. Since a spin wave is composed
of an electron and a hole, it is, indeed, neutral.
At a formal level, the creation operator for an
${S_z}=1$ spin wave,
\begin{eqnarray}
{S_+}({\bf x},t) = {c^\dagger_\uparrow}({\bf x},t)
\,{c_\downarrow}({\bf x},t)
\end{eqnarray}
is invariant under a gauge transformation,
${c_\alpha}({\bf x},t)\rightarrow
{e^{i\phi({\bf x},t)}}{c_\alpha}({\bf x},t)$.
Consequently, such an operator does not couple to
the electromagnetic field through minimal coupling.

Nevertheless, a spin wave {\em does} carry current.
When the vector potential, ${\bf A}$ vanishes,
the current takes the form
\begin{equation}
{\bf J} = {\sum_{\bf k}} \frac{e}{m}\,{\bf k}\,
{c^\dagger_\alpha}({\bf k})\,{c_\alpha}({\bf k})
\label{eqn:current-op}
\end{equation}
The current operator has this form irrespective of
the electron-electron interaction terms, 
so long as they are Galilean-invariant --
i.e. so long as they are momentum-independent and
translationally-invariant.

Consider the operator which creates a spin wave of momentum
${\bf q}$:
\begin{eqnarray}
{S_+}({\bf q}) = {\sum_{\bf k}} {c^\dagger_\uparrow}({\bf k}+{\bf q})
\,{c_\downarrow}({\bf k})
\end{eqnarray}
In so doing, it actually creates current
as well, as may be seen by taking its commutator
with the current operator
\begin{equation}
\left[{\bf J},{S_+}({\bf q})\right] =
\frac{e}{m}\,{\bf q}\,{S_+}({\bf q})
\end{equation}

Hence, {\rm spin waves carry current}. This is a
purely kinematic statement which follows from the
form of the current operator (\ref{eqn:current-op})
which, in turn, follows from Galilean invariance.
Our conclusion holds whether or not the electron liquid
orders electronically.

However, it may be difficult to see how
this electrical current appears in an effective field theory
of spin waves in, for instance, the ferromagnetic state.
Suppose we take our Galilean-invariant electronic
Lagrangian,
\begin{equation}
{\cal L} = {c^\dagger_\alpha}\left(i{\partial_t}-e{A_t}\right){c_\alpha}
+ \frac{1}{2m}\,{c^\dagger_\alpha}
{\left(i{\bf \nabla}-e{\bf A}\right)^2}{c_\alpha} \:+\:
{{\cal L}_{\rm int}}
\end{equation}
and decouple ${{\cal L}_{\rm int}}$ with a Hubbard-Stratonovich field
${\bf S}$ which couples linearly to
${c^\dagger_\alpha}~{{\bf \sigma}_{\alpha\beta}}~{c_\beta}$.
We can integrate out the electrons and expand the resulting
action about a ferromagnetic state which is ordered in
the ${\bf \hat z}$ direction. On general grounds,
we expect that the resulting effective action
will be of the form
\begin{equation}
{{\cal L}_{\rm eff}} = {S_+} \,i{\partial_t} {S_-}
- D\,{\bf \nabla}{S_+}\cdot{\bf \nabla}{S_-} \:+\:
\ldots
\label{eqn:ferr-action}
\end{equation}
As we noted above, $S_\pm$ is invariant under
a gauge transformation, so it is hard to imagine how
it can be coupled to the electromagnetic field, ${\bf A}$.
On the other hand, ${\bf J}={\partial}{\cal L}/{\partial}{\bf A}$,
so there will be no current carried by $S_\pm$
in the absence of such a coupling.

\begin{figure*}[h]
\epsfysize=7.5cm 
\centerline{\epsfbox{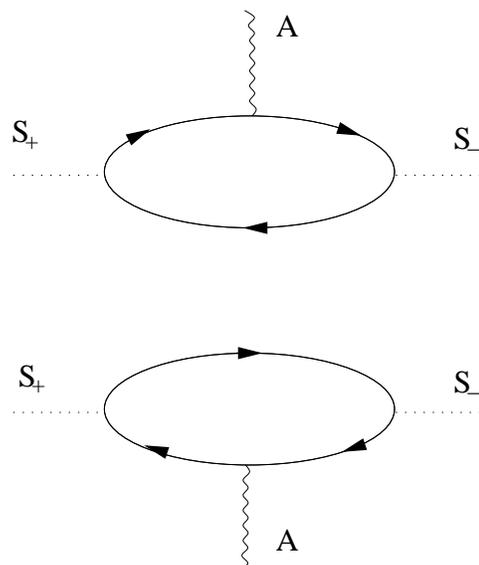}}
\caption{The diagrams which contribute to the coupling between
spin waves and the electromagnetic field.}
\label{fig:coup-diag}
\end{figure*}

The resolution is that there is a coupling to ${\bf A}$
hidden in the ``$\ldots$'' in (\ref{eqn:ferr-action}). If it is
difficult to guess the form of this term, it is
because we would be wrong in assuming that it is local.
Since we have integrated out gapless fermionic degrees
of freedom in obtaining (\ref{eqn:ferr-action}), we should
actually expect non-local terms. There are no non-local
terms in the spin dynamics of (\ref{eqn:ferr-action}) because
the up- and down-spin Fermi wavevectors are different
as a result of the development of ferromagnetic order;
consequently spinful excitations of the Fermi surface
have a minimum wavevector. However, the charged
excitations extend down to ${\bf q}=0$,
and the coupling of $S_\pm$ to ${\bf A}$
is, indeed, non-local. It may be obtained by computing
the diagrams of Fig. \ref{fig:coup-diag} and takes the form:
\begin{equation}
{{\cal L}_{\rm A}} = \frac{e}{m}\,
{{\bf A}^T}\cdot{S_+}i{\bf \nabla}{S_-}
\label{eqn:coup-action}
\end{equation}
In this equation, ${{\bf A}^T}$ denotes the
transverse part of ${\bf A}$, which is given in 
momentum space by:
\begin{equation}
{{\bf A}^T}({\bf q}) = {\bf A}({\bf q}) - {\bf q}\,
\frac{{\bf q}\cdot{\bf A}({\bf q})}{q^2}
\end{equation}
This is both non-local and {\em gauge-invariant} since,
a gauge transformation,
\begin{equation}
{\bf A}({\bf q}) \rightarrow {\bf A}({\bf q}) +
{\bf q}\,\phi({\bf q})
\end{equation}
with $\phi({\bf q})$ arbitrary, leaves ${{\bf A}^T}({\bf q})$
unchanged. Since ${S_+}{\bf \nabla}{S_-}$
is also invariant under a gauge transformation, the
entire term (\ref{eqn:coup-action}) is gauge-invariant,
which is a cause for some relief.

Note that spin waves were empowered with the ability
to carry a current by the gapless charge degrees of freedom
with which they interact. In an insulating ferromagnet,
spin waves will not carry a current proportional to their
momentum. Since insulating behavior will only occur when
a system is not translationally-invariant, there is no
contradiction here.

\section{Quasiparticles in a Superconductor}

The Bogoliubov-de Gennes quasiparticles of a superconductor
are coherent superpositions of electrons and holes. Hence,
they do not have a well-defined charge. As the Fermi surface is
approached, a Bogoliubov-de Gennes quasiparticle
becomes an equal superposition of electron and hole; thus,
one might be tempted to assign it zero charge in
this limit. This is not an academic question in
an unconventional superconductor such as one
of $d_{{x^2}-{y^2}}$ symmetry -- as the high-$T_c$ cuprates
are believed to be -- since, in the absence of a
full gap, quasiparticles will be thermally
excited down to zero temperature and their ability to carry current
will have an impact on the superfluid density.

For the sake of concreteness, let us consider a two-dimensional
$d_{{x^2}-{y^2}}$ superconductor and focus on its
nodal quasiparticles. We assume that the system is
Galilean-invariant so that the order parameter
spontaneously breaks rotational symmetry
when it chooses nodal directions. The effective action
for a superconductor is of the form
\begin{eqnarray}
S &=& \int_{{k_y}>0} \frac{{d^2}k}{(2\pi)^2}\,dt\:
{\Psi^\dagger}(k,t) \biggl[ \left(i{\partial_t}-{\tau^z}e{A_t}\right)\cr
& & {\hskip 2 cm}
- {\tau^z}\left(\epsilon\left({\bf k}+{\tau^z}e{\bf A}\right)-\mu\right)\cr
& & {\hskip 2 cm}
- {\tau^+} \Delta(k) - {\tau^-} {\Delta^\dagger}(k)\biggr]{\Psi}(k,t)
\end{eqnarray}
where we have used the Nambu-Gorkov notation:
\begin{equation}
\Psi_{a\alpha}(\vec{k})  = \left[ \begin{array}{c}
c_{k\uparrow}^{\vphantom\dagger} \\ c_{-k\downarrow}^{\dagger} \\ 
c_{k\downarrow}^{\vphantom\dagger} \\ -c_{-k\uparrow}^{\dagger} 
\end{array}\right].
\end{equation}
and the $\tau^i$ are Pauli matrices which act on the particle-hole
index $a$. If we consider the four-component object as composed
of two two-component blocks, the upper and lower blocks,
then the $\tau^i$ mix the components within a block. There
are also Pauli matrices $\sigma^i$ which act on the spin
indices, $\alpha$, and mix the upper block with the lower
block.

We will linearize this action about the nodes of
$\Delta(k)={\Delta_0}(\cos{k_x}a - \cos{k_y}a)$.
We must retain two fermion fields, one for each
pair of antipodal nodes, but these pairs
of nodes are not coupled to each
other in the low-energy
limit, so we will often focus on just one.
By linearizing about the nodes, we are approximating
the momentum of an electron by $k_F$ and discarding the
deviation from the Fermi surface. Hence, we will
verify that the relation ${\bf J} = \frac{e}{m}\,{\bf P}$ is
satisfied {\em to this level of approximation},
which means ${\bf J} = \frac{e}{m}\,{N_{\rm qp}}\,{{\bf k}_F}$,
where $N_{qp} = {\Psi^\dagger} \Psi$ is the difference
between the number of electrons at one node and the
number at the antipodal node.
If we kept the full Galilean-invariant
expression $\epsilon(k)={k^2}/2m$, then we could
verify ${\bf J} = \frac{e}{m}\,{\bf P}$ exactly. We
will do this in one-dimension, where it is particularly
instructive. For now, we will content ourselves with
a crude verification.

We align our coordinate system along the nodal
dierection and linearize the single-particle dispersion:
$\epsilon(k)~-~\mu~\approx~\frac{k_F}{m}~{k_x}$,
where $k_x$ is the momentum perpendicular to the
Fermi surface, measured away from the node. A similar
expression holds for the other pair of nodes,
with $k_x$ replaced by $k_y$. We also
linearize the gap about the nodes;
\begin{equation}
\Delta \tau^+  \approx v_{\scriptscriptstyle\Delta} 
{\tau^+} e^{ie\varphi/2} \left(-i{\partial_y}\right)
e^{ie\varphi/2}
\end{equation}
where $e^{ie\varphi}$ is the phase of the superconducting
order parameter. Some care was needed in obtaining the
correct ordering of derivatives and $\varphi$s; for
details, see refs. \onlinecite{Balents98,Yang01}.
Integrating out the electronic states far from
the nodes and the fluctuations of the
amplitude of the order parameter, we obtain the action
\begin{eqnarray}
S &=& \int 
{\Psi^\dagger} \biggl[ \left(i{\partial_t}-{\tau^z}e{A_t}\right)
+ {\tau^z}\frac{k_F}{m}\left(i{\partial_x}-{\tau^z}e{A_x}\right)\cr
& & {\hskip 1.2 cm}
+\: {v_{\scriptscriptstyle\Delta}}{\tau^s}
{e^{ies\varphi/2}}i{\partial_y}{e^{ies\varphi/2}}
\biggr]{\Psi}\cr
& & +\:\frac{1}{2}\,{\rho_s}\int \left[\frac{1}{v_c^2}
{\left({\partial_t}\varphi+2{A_t}\right)^2}
- {\left({\partial_i}\varphi+2{A_i}\right)^2}\right]\cr
& & +\:\:\ldots
\end{eqnarray}
where $s=\pm$, and $\rho_s$ and $v_c$ are the bare superfluid
density and velocity.
The ``\ldots'' includes the action for 
the other pair of nodes and higher-order terms, which we neglect.

Following Ref. \onlinecite{Balents98}, we can simplify this
action by defining {\em neutral quasiparticles},
$\chi$, according to \cite{multi-valued}:
\begin{equation}
\chi = \exp(-ie\varphi \tau^z/2) \Psi  .
  \label{transform}
\end{equation}
The action now takes the form:
\begin{eqnarray}
\label{eqn:neutral-action}
S &=& \int{\chi^\dagger}\biggl[ i{\partial_t}
+ {\tau^z}\frac{k_F}{m}i{\partial_x}
+ {v_{\scriptscriptstyle\Delta}}{\tau^x} i{\partial_y}
\biggr]{\chi}\cr
& & -\: \frac{1}{2}\,\int\biggl[ e{\chi^\dagger}{\tau^z}{\chi}
\left({\partial_t}\varphi+2{A_t}\right)\cr
& & {\hskip 1.2 cm}
+ e{\chi^\dagger}{\chi}
\frac{k_F}{m}\left({\partial_x}\varphi+2{A_x}\right)\biggr] \cr
& & +\:\frac{1}{2}\,{\rho_s}\int \left[\frac{1}{v_c^2}
{\left({\partial_t}\varphi+2{A_t}\right)^2}
- {\left({\partial_i}\varphi+2{A_i}\right)^2}\right]\cr
& & + \:\: \ldots
\end{eqnarray}
The quasiparticle annihilation operator, $\chi$,
is gauge-invariant since it is neutral, but $\varphi$,
which is charged, is not.
The action (\ref{eqn:neutral-action})
is gauge-invariant because $\chi$ is only coupled
to gauge-invariant quantities,
such as the superfluid density and current,
${\partial_\mu}\varphi+2{A_\mu}$.

These neutral excitations nevertheless carry current.
By differentiating the Lagrangian of (\ref{eqn:neutral-action})
with respect to $A_x$, we find that the current in the $x$-direction is
\begin{equation}
{J_x} = 2{\rho_s}\left({\partial_x}\varphi + 2{A_x}\right)
+ \frac{e}{m}\,{k_F}\,{\chi^\dagger}{\chi}
\end{equation}
The first term is the supercurrent; it derives from the
final line of (\ref{eqn:neutral-action}). The second term comes from
the third line of (\ref{eqn:neutral-action}), and it states
that the quasiparticles carry a current which is
$e/m$ times their momentum $k_F$:
\begin{equation}
{J_x^{\rm qp}} = \frac{e}{m}\,{k_F}\,{\chi^\dagger}{\chi}
\end{equation}
By differentiating (\ref{eqn:neutral-action}) with
respect to $A_t$, we find that the corresponding
charge density is
$\rho=-2({\rho_s}/{v_c^2})
\left({\partial_t}\varphi~+~2{A_t}\right)+
e{\chi^\dagger}{\tau^z}{\chi}$. The second term is the
quasiparticle contribution.
Although the quasiparticles are neutral in the sense
of being gauge-invariant, they contribute to both the
charge and current densities.

Suppose that we integrate out the fluctuations
of the phase of the superconducting order
parameter. What does the coupling between the
quasiparticles and the electromagnetic field look like?

To integrate out $\varphi$, it is convenient to
use the dual representation in which $\varphi$
is replaced by a dual gauge field, $a_\mu$. In
this dual representation, (\ref{eqn:neutral-action})
takes the form (for details, see ref. \onlinecite{Balents98}).
\begin{eqnarray}
\label{eqn:dual-action1}
S &=& \int{\chi^\dagger}\biggl[ i{\partial_t}
+ {\tau^z}\frac{k_F}{m}i{\partial_x}
+ {v_{\scriptscriptstyle\Delta}}{\tau^x} i{\partial_y}
\biggr]{\chi}\cr
& & {\hskip - 1 cm}-\:\int  \frac{1}{2\rho_s}\,
{\left({\epsilon_{\mu\nu\lambda}}
{\partial_\nu}{a_\lambda}\right)^2} +
\left(2{A_\mu}-\frac{1}{2\rho_s}{J_\mu^{\rm qp}}\right)
{\epsilon_{\mu\nu\lambda}}{\partial_\nu}{a_\lambda}\cr
& &{\hskip - 1 cm} +\:\: \ldots
\end{eqnarray}
where we have chosen units with ${v_c}=1$ to facilitate
the use of `relativistic' notation.
The dual gauge field $a_\mu$ is related to the total current,
$J_\mu$ by
\begin{equation}
{J_\mu} = {\epsilon_{\mu\nu\lambda}}{\partial_\nu}{a_\lambda}
\end{equation}
It only enters the action in this transverse combination
which is automatically conserved. Furthermore, this
means that ${\epsilon_{\mu\nu\lambda}}{\partial_\nu}{a_\lambda}$
is only coupled to the transverse
parts of ${A_\mu}$ and ${J_\mu^{\rm qp}}$.
Since it appears quadratically, we can now integrate
it out, obtaining:
\begin{eqnarray}
S &=& \int{\chi^\dagger}\biggl[ i{\partial_t}
+ {\tau^z}\frac{k_F}{m}i{\partial_x}
+ {v_{\scriptscriptstyle\Delta}}{\tau^x} i{\partial_y}
\biggr]{\chi}\cr
& & - \int {A^T_\mu} {J_\mu^{T\:{\rm qp}}}\:\:
 + \:\: \ldots
\end{eqnarray}
The coupling between $A_\mu$ and
${J_\mu^{\rm qp}}$ is non-local because it
only couples their transverse parts,
as in (\ref{eqn:coup-action}).
Again, since we have integrated out $a_\mu$ which is
formally a gapless degree of freedom when $A_\mu$
is held fixed, we should not be surprised by the appearance
of a non-local coupling between $A_\mu$ and
${J_\mu^{\rm qp}}$ through which only their
gauge-invariant transverse components are coupled.
Since $a_\mu$ does not couple to the longitudinal parts
of $A_\mu$ and ${J_\mu^{\rm qp}}$, it is not possible to
generate terms involving them.

Again, the ability of quasiparticles to carry a current
depends on their interaction with gapless charged
degrees of freedom -- in this case, a supercurrent. If
the conductivity associated with this supercurrent
(i.e. its Drude weight, {\em not} its superfluid density
\cite{Scalapino92}, see section V) is reduced, e.g. by
the localization of some electrons at impurities, then
Bogoliubov-DeGennes quasiparticles will carry a reduced
current as well.

\section{Spin-Charge Separated One-Dimensional Electron Gas}

The one-dimensional electron gas can be described
completely in terms of its spin- and charge- collective
modes \cite{Emery79}. The electron itself is a combination of
charge- and spin-carrying solitons -- holons and spinons --
in these collective modes. Because these collective
modes have different
velocities, the charge and spin of an electron move
apart in time. Both the charge and spin modes
can carry momentum, but one might assume that only the
charged mode should couple to the electromagnetic
field and carry current. Furthermore, the velocities $v_c$
and $v_s$ of these modes depend on the interaction strength;
they are, in general, different from ${k_F}/m$, which
might lead one to expect that even the charged mode will
carry a current which is not equal to $\frac{e}{m}$ times
its momentum. However, we have come,
by now, to distrust such expectations.

The Hamiltonian density is often written in the bosonized form
\begin{eqnarray}
\label{eqn:naive-bose-ham}
{\cal H} &=& \frac{1}{2}\,{v_c}\left[
{K_c}\, {\left({\partial_x}{\varphi_c}\right)^2}
\,+\, \frac{1}{K_c}\,{\left({\partial_x}{\theta_c}\right)^2}
\right]\cr
& &+\:\frac{1}{2}\,{v_s}\left[
{K_s}\, {\left({\partial_x}{\varphi_s}\right)^2}
\,+\, \frac{1}{K_s}\,{\left({\partial_x}{\theta_s}\right)^2}
\right]\cr
& & +\: {v_F}{k_F}\sqrt{\frac{2}{\pi}}\,{\partial_x}{\theta_c}
\end{eqnarray}
The final term is the Fermi energy (for a perfectly linear
spectrum) multiplied by the electron number. This
term is cancelled by the chemical potential, but
we have retained it for purposes of comparison
with the corresponding expression for the
momentum density.
If the system respects $SU(2)$ spin-rotational symmetry, then
${K_s}=1$. If ${K_c}=1$ as well, then the Hamiltonian
describes free fermions. As $K_c$ is shifted away from
$1$ by the interactions, the charge of the fundamental
charged soliton is also shifted away from $e$.
$\varphi_c$ and $\theta_c$ are dual variables,
${v_c}{\partial_x}{\theta_c}={K_c}{\partial_t}{\varphi_c}$,
as are $\varphi_s$ and $\theta_s$. They are symmetric
and anti-symmetric combinations of left- and right-moving
fields, ${\theta_c}={\phi_{cR}}-{\phi_{cL}}$,
${\varphi_c}={\phi_{cR}}+{\phi_{cL}}$.
The charge and spin modes are symmetric and anti-symmetric
combinations of up- and down-spin modes,
${\theta_c}=({\theta_\uparrow}+{\theta_\downarrow})/\sqrt{2}$,
${\theta_s}=({\theta_\uparrow}-{\theta_\downarrow})/\sqrt{2}$,
etc.

This Hamiltonian describes the physics of interacting
fermions with a spectrum which is linearized about
the Fermi surface, $\pm k_F$. The annihilation operator
for a right-moving spin-up electron is
\begin{equation}
{\psi_{R\uparrow}} = \frac{1}{\sqrt{2\pi a}}\,
{e^{-i\sqrt{\frac{\pi}{2}}\left({\varphi_c}+{\theta_c}\right)}}
\:{e^{-i\sqrt{\frac{\pi}{2}}\left({\varphi_s}+{\theta_s}\right)}}
\end{equation}
where $a$ is a short-distance cutoff.
Similar relations hold for down-spin, right-moving electrons
and left-moving electrons of both spins.
The right- and left-moving charge densities are:
\begin{eqnarray}
\rho_{R,L} = \frac{1}{\sqrt{2\pi}}\,
{\partial_x}\left({\theta_c}\pm{\varphi_c}\right)
\end{eqnarray}
The right- and left-moving $S_z$ densities
are given by a similar expression with
${\theta_c}$,${\varphi_c}$ replaced by
${\theta_s}$,${\varphi_s}$.

The momentum can be obtained from the energy-momentum
tensor, $T_{\mu\nu}$. While the Hamiltonian density is the $tt$
component, ${\cal H} = {T_{tt}}$, the momentum density is
given by $P = {T_{tx}}$.
\begin{equation}
\label{eqn:bose-momentum}
P = {k_F}\sqrt{\frac{2}{\pi}}\,{\partial_x}{\varphi_c} + 
\,\left[
\left({\partial_x}{\varphi_c}\right)\left({\partial_x}{\theta_c}\right)
\:+\: \left({\partial_x}{\varphi_s}\right)
\left({\partial_x}{\theta_s}\right)\right]
\end{equation}
Note that this takes a somewhat
different form than is usual for relativistic scalar
fields since excitations about the ground state
are centered at $\pm {k_F}$; the first term would not be
present in an ordinary relativistic
system at zero-density, where low-energy excitations
are centered about $k=0$. It is the counterpart
to the final term in (\ref{eqn:naive-bose-ham}); it assigns momentum
$\pm k_F$ to each right- or left-mover. The second term
accounts for possible changes in the local value
of $k_F$.

In order to determine the current operator, we
modify the Hamiltonian via minimal
coupling, which replaces ${\partial_x}{\varphi_c}$ with 
${\partial_x}{\varphi_c}-e\sqrt{\frac{2}{\pi}}{A_x}$.
We now differentiate
with respect to $A_x$ to obtain ${J_x}=-\partial{\cal H}/\partial{A_x}$.
This coupling is dictated by the
fact that ${\varphi_c}\rightarrow{\varphi_c}-e\sqrt{\frac{2}{\pi}}\chi$
when $\psi_{R,L\:\alpha}\rightarrow
{e^{ie\chi}}\psi_{R,L\:\alpha}$,
${A_x}\rightarrow{A_x}-{\partial_x}\chi$.
Since ${v_c}{\partial_x}{\theta_c}={K_c}{\partial_t}{\varphi_c}$,
it does not couple to $A_x$.

However, before we do this, we need to exercise some
care with regards to Galilean invariance.
We would like to consider only momentum-independent
interactions. Hence, the interaction terms
cannot have independent coefficients $\lambda_{RR}$
and $\lambda_{RL}$ for the
${\rho_R}\,{\rho_R} + {\rho_L}\,{\rho_L}$
interaction and the ${\rho_R}\,{\rho_L}$ interaction.
The only allowed local interaction between charge densities
is a simple density-density interaction of the form
\begin{eqnarray}
\lambda \,\rho\,\rho &=& \lambda\,{({\rho_R} + {\rho_L})^2}\cr
&=& \lambda\left({\rho_R}\,{\rho_R} + {\rho_L}\,{\rho_L}\right)
+ 2\lambda\,{\rho_R}\,{\rho_L}\cr
&=& \frac{2\lambda}{\pi} \, {\left({\partial_x}{\theta_c}\right)^2}
\end{eqnarray}
i.e. $\lambda_{RL} = 2 \lambda_{RR}$.
If $\lambda_{RL} \neq 2 \lambda_{RR}$,
the Hamiltonian will contain a term
of the form $({\rho_R} - {\rho_L})^2$,
which is proportional to the total momentum
squared, in which case the Hamiltonian is
not Galilean-invariant. This is the case for
the edge states of a quantum Hall bar or
quantum Hall line junction \cite{Mitra01}.
Hence, when we look at the charged sector
of the Hamiltonian (\ref{eqn:naive-bose-ham}),
which arises by combining the free and interaction
terms,
\begin{eqnarray}
{{\cal H}_{\rm charge}} &=&  \frac{1}{2}\,\frac{k_F}{m}\left[
{\left({\partial_x}{\varphi_c}\right)^2}
+ \,{\left({\partial_x}{\theta_c}\right)^2}
\right] +  \frac{2\lambda}{\pi} \, {\left({\partial_x}{\theta_c}\right)^2}\cr
&=& \frac{1}{2}\,{v_c}\left[{K_c}\,
{\left({\partial_x}{\varphi_c}\right)^2}
\,+ \,\frac{1}{K_c}\,{\left({\partial_x}{\theta_c}\right)^2}
\right]
\end{eqnarray}
we see that ${v_c}{K_c}={k_F}/m$. In other words,
in a Galilean-invariant system, the change in the
charge velocity is precisely compensated by the
change in the soliton charge so that their product,
which will determine the current, is the same as
the free fermion value, ${k_F}/m$.

The second point which requires some care
is the linearization of the Hamiltonian.
By linearizing our Hamiltonian about the Fermi surface,
we are approximating our system by a `relativistic'
one. In a relativistic system, the current density
and momentum density cannot be proportional to each other
since the former is the spatial component of
a vector, $J_\mu$, and the other is a component of
a tensor $T_{\mu\nu}$ (the total momentum {\em is}
the spatial component of vector, but this is obtained
by integrating the momentum density over
the entire system); a relation of the form
${J_x}=\frac{e}{m}{T_{tx}}$ would break `relativistic'
invariance. Hence, we need to retain the terms which
break `relativistic' invariance and contain the
information about Galilean invariance. While the linearized
terms in the Hamiltonian are of the form ${v_F}(k-{k_F})$,
the terms which `know' about Galilean invariance
are of the form ${(k-{k_F})^2}/2m$. These terms actually
couple the spin and charge modes, thereby resulting in an
electrical current carried by spinons.

To see this, consider a term in the Hamiltonian
which gives a quadratic spectrum, ${(k-{k_F})^2}/2m$,
and its bosonized form:
\begin{eqnarray}
{\psi^\dagger_{R\uparrow}}\,\frac{1}{2m}\,{\left({i\partial_x}\right)^2}
{\psi_{R\uparrow}} &=& \cr
& & {\hskip - 2 cm}\frac{1}{3}\,\frac{1}{2\pi}\,\frac{1}{2m}\,
{\left[{\partial_x}
\sqrt{\frac{\pi}{2}}\left({\varphi_c}+{\theta_c}+{\varphi_s}+{\theta_s}\right)
\right]^3}\cr
& &+ \: \text{total derivative terms}
\end{eqnarray}
Hence, summing over both spins and over right- and left-movers,
we have
\begin{eqnarray}
{\psi^\dagger_{R\alpha}}\,\frac{1}{2m}\,{\partial_x^2}
{\psi_{R\alpha}} \: + \: {\psi^\dagger_{L\alpha}}\,
\frac{1}{2m}\,{\partial_x^2}
{\psi_{L\alpha}}\: = {\hskip  2 cm}\cr
\frac{1}{2m}\,\sqrt{\frac{\pi}{2}}\,\left[
{\left({\partial_x}{\varphi_c}\right)^2}\left({\partial_x}{\theta_c}\right)
\:+\: 2\left({\partial_x}{\varphi_c}\right)
\left({\partial_x}{\varphi_s}\right)
\left({\partial_x}{\theta_s}\right)\right]\cr
+ \: \text{terms which do not contain $\varphi_c$}
\end{eqnarray}
In a Galilean-invariant system, with single-particle
kinetic energy ${k^2}/2m$, these are the only other
terms which we must add.

Hence, going beyond linearization about the Fermi
points and retaining the quadratic single-particle
spectrum of a Galilean-invariant system, we have
the following Hamiltonian:
\begin{eqnarray}
{\cal H} &=& \frac{1}{2}\,\left[
\frac{k_F}{m}\, {\left({\partial_x}{\varphi_c}\right)^2}
\,+\, \frac{v_c}{K_c}\,{\left({\partial_x}{\theta_c}\right)^2}
\right]\cr
& &+\:\frac{1}{2}\,{v_s}\left[
{K_s}\, {\left({\partial_x}{\varphi_s}\right)^2}
\,+\, \frac{1}{K_s}\,{\left({\partial_x}{\theta_s}\right)^2}
\right]\cr
& & {\hskip - 1 cm}+\, \frac{1}{2m}\,\sqrt{\frac{\pi}{2}}\,\left[
{\left({\partial_x}{\varphi_c}\right)^2}\left({\partial_x}{\theta_c}\right)
\:+\: 2\left({\partial_x}{\varphi_c}\right)
\left({\partial_x}{\varphi_s}\right)
\left({\partial_x}{\theta_s}\right)\right]\cr
& & + \: \text{terms which do not contain $\varphi_c$}
\end{eqnarray}

If we now apply minimal coupling,
${\partial_x}{\varphi_c}\rightarrow{\partial_x}{\varphi_c}~-~e
\sqrt{\frac{2}{\pi}}{A_x}$ and differentiate
with respect to $A_x$ to obtain ${J_x}=-\partial{\cal H}/\partial{A_x}$,
we find the current operator:
\begin{equation}
{J_x} = e\frac{k_F}{m}\sqrt{\frac{2}{\pi}}{\partial_x}{\varphi_c} + 
\frac{e}{m}\,\left[
\left({\partial_x}{\varphi_c}\right)\left({\partial_x}{\theta_c}\right)
\:+\: \left({\partial_x}{\varphi_s}\right)
\left({\partial_x}{\theta_s}\right)\right]
\end{equation}
Comparing this expression with
(\ref{eqn:bose-momentum}), we see that it satisfies
the relation $J=\frac{e}{m}P$.

The charged field, ${\partial_x}\varphi_c$, carries large momentum
${k_F}$. The spin field, ${\partial_x}\varphi_s$, only carries
the small momentum of deviations from the Fermi points
(as do ${\partial_x}\theta_c$ and ${\partial_x}\theta_s$).
Hence, the latter can get lost in the shuffle if we
only keep the leading terms in a gradient expansion
about the Fermi points. To see the relation between
current and momentum, we must keep the quadratic terms in both.

Note that the condition  $\lambda_{RL} = 2 \lambda_{RR}$
as well as the constraints on the cubic terms in the
Hamiltonian both followed from Galilean invariance.
Even a mild breaking of this invariance such as that
caused by a lattice which is far from any nesting
condition could lead to a violation of these
conditions and, hence, of the relation between the current and
the momentum.

\section{Spin-Charge Separation in $2+1$-Dimensions}

We will describe a spin-charge separated state
as a quantum disordered superconducting state
\cite{Balents98,Balents99}
which has the advantage, in the current context,
of allowing us to take as our starting point
the discussion in section III of Bogoliubov-DeGennes
quasiparticles in a superconductor.

A quantum-disordered $d$-wave superconductor can be described by
an extension of (\ref{eqn:dual-action1}) to include vortices:
\begin{eqnarray}
\label{eqn:dual-action2}
S &=& \int{\chi^\dagger}\biggl[ i{\partial_t}
+ {\tau^z}\frac{k_F}{m}i{\partial_x}
+ {v_{\scriptscriptstyle\Delta}}{\tau^x} i{\partial_y}
\biggr]{\chi}\cr
& & -\:\int  \frac{1}{2\rho_s}{\left({\epsilon_{\mu\nu\lambda}}
{\partial_\nu}{a_\lambda}\right)^2} +
\left(2{A_\mu}-\frac{1}{2\rho_s}{J_\mu^{\rm qp}}\right)
{\epsilon_{\mu\nu\lambda}}{\partial_\nu}{a_\lambda}\cr
& & +\:\int {\left|\left(i{\partial_\mu} + 2{a_\mu}\right)
{\Phi_{hc/e}} \right|^2} - V\!\left({\Phi_{hc/e}}\right)
\end{eqnarray}
The last line of (\ref{eqn:dual-action2}) implements
the Magnus force interaction between vortices
and the supercurrent.
Here, ${\Phi_{hc/e}}$ is the annihilation operator for
a flux $hc/e$ vortex, which we assume is the lightest
vortex near the quantum-disordered state. If
$hc/2e$ vortices condense instead, then
spin and charge are confined \cite{Balents99}.
When superconductivity is destroyed through the
condensation of flux $hc/e$ vortices, the resulting
state supports `holons', which are spinless, charge-$e$ solitons in
the vortex condensate, and `nodons' or `spinons' $\chi$, which are
`neutral' spin-$1/2$ excitations.

When vortices condense, superconductivity is
destroyed because magnetic flux is no longer expelled.
In other words, the superfluid density vanishes.
In the dual description offered in (\ref{eqn:dual-action2}),
charges ${\epsilon_{ij}}{\partial_i}{a_j}$ enter the
vortex condensate in a lattice -- a `holon Wigner crystal'
\cite{Balents98,Balents99}.
However, in a Galilean-invariant system, the Wigner
crystal can slide. Hence, even though the superconductivity
is destroyed with the disappearance of the Meissner effect,
the system is still a perfect conductor. Thus,
when we integrate out $a_\mu$, we will still obtain
a coupling between $A_\mu$ and ${J_\mu^{\rm spinon}}$
which is of the form
\begin{equation}
\int {A_\mu}\,\left\langle {\epsilon_{\mu\alpha\beta}}
{\partial_\alpha}{a_\beta}
{\epsilon_{\nu\gamma\delta}}
{\partial_\gamma}{a_\delta}
\right\rangle\,{J_\nu^{\rm spinon}}
\end{equation}
In the limit of ${\bf q}=0$, $\omega\rightarrow 0$,
this is determined by the conductivity -- or Drude weight --
which is the same as in the superconducting case.
(It is not determined by the Meissner or diamagnetic response,
which vanishes.) Hence, upon integrating out $a_\mu$ and
${\Phi_{hc/e}}$, we obtain the same induced coupling between
spinons and the electromagnetic field,
${A^T_\mu} {J_\mu^{T\:{\rm spinon}}}$, that we obtained
for Bogoliubov-de Gennes quasiparticles.

However, even {\em infinitesimal} translational-symmetry
breaking, such as that caused by a small density
of impurities, will pin the holon Wigner crystal. Consequently,
the system will be an insulator and $a_\mu$ will be
gapped. The coupling between spinons and the 
electromagnetic field will now be of the form \cite{Balents98},
\begin{equation}
{S_{\rm coupling}} = \int {A_\mu}\left({\partial^2}{J_\mu^{{\rm spinon}}}
- {\partial_\mu}{\partial_\nu}{J_\nu^{{\rm spinon}}}\right)
\end{equation}
In other words, spinons will be truly neutral since
they do not carry a current proportional to their
momentum density, in contrast to the
merely `neutral' spinons that do.  

Note that holons are not necessarily bosonic.
A bosonic holon can form a bound state with
an uncondensed $hc/2e$ vortex, or `vison',
thereby becoming fermionic \cite{Demler01}. In this case,
the holon Wigner crystal is not the only possible
non-superconducting ground state because the
holons could form a perfectly conducting Fermi liquid.
If the spinons pair and form a spin
gap, then a spin-gapped metallic state can result,
in which spinons carry a current proportional to their
momentum density.

\section{Discussion}

The basic form of the interaction between
`neutral' and `charged' quasiparticles
is ${J_\mu^{\rm neutral}}\,{J_\mu^{\rm charged}}$.
It can be interpreted as a `Doppler shift'
by which the motion of the neutral quasiparticles
brings the charged ones along for the ride.
As a result, the relation
${\bf J}({\bf x}) = \frac{e}{m} {\bf P}({\bf x})$
is satisfied even in a system with formally
`neutral' quasiparticles.
The facility with which the charge carriers can
move along with the neutral quasiparticles is,
of course, a consequence of Galilean invariance.

Even a mild violation of Galilean invariance
can have dramatic consequences for the
relationship between current and momentum
and, hence, for the conductivity. In the case of
spin-charge separation in $2+1$ dimensions,
we saw that infinitesimal translation symmetry
breaking can make a perfect conductor
into an insulator; as a consquence, `neutral' quasiparticles
which carry a current proportional to their momentum
become truly neutral quasiparticles carrying no current.
Similarly, spin waves in a Galilean-invariant electron
system carry current, but spin waves in an insulating ferromagnet
on a lattice do not carry current. Thus the lattice
has a large effect on the electrical properties
of spin waves, even though it
does not seem to be particularly important for
the magnetic properties of the ferromagnet phase.
Even in $1+1$ dimensions, in those situations in which
the effects of the ionic lattice are otherwise mild because the Fermi surface
is far from nested, the relation between current
and momentum can be strongly violated as a result of
the effect of the lattice on interaction parameters
and `small' corrections to the band dispersion.

We wish to thank S. Kivelson and T. Senthil for helpful
discussions. C. N. was supported by the
National Science Foundation under
Grant No. DMR-9983544 and the A.P. Sloan Foundation.
K.S. was supported by the Department of
Energy under grant No. DE-FG03-00ER45843
S.M.G. was supported by the National Science Foundation under
Grant No. DMR-0087133 and PHY9907949. M.P.A.F. was supported by
the National Science Foundation under
Grant No. DMR9704005 and PHY9907949.

%\bibliography{../../bibs/corr}

\begin{thebibliography}{99}
\bibitem{multi-valued} The field $\chi$ is multi-valued in the
presence of a flux $hc/2e$ vortex, about which $\varphi$ winds by $2\pi$.
This can be incorporated by introducing Chern-Simons
gauge fields which mediate the interaction
between vortices and quasiparticles \cite{Balents99}.
Since vortices also interact with
the total electrical current through the Magnus force, they can mediate
an interaction between the $\chi$s and the electrical
current. For simplicity, we will assume that there are no
vortices in our superconductor. When we consider spin-charge
separation, we will assume that flux $hc/2e$ vortices
remain gapped and assume that $hc/e$ vortices
condense. The field $\chi$ is single-valued
in the presence of flux $hc/e$ vortices, so
there is no problem.
\end{thebibliography}
%\bibliographystyle{prsty}

\end{document}